\documentclass[journal]{IEEEtran}

\ifCLASSINFOpdf
\else
   \usepackage[dvips]{graphicx}
\fi
\usepackage{url}
\usepackage{graphicx}
\usepackage{caption}
\usepackage{afterpage}
\usepackage{amsmath,amssymb,amsfonts, mathtools}%
\usepackage{makecell}
\usepackage{bm}
\usepackage{upgreek}
\usepackage{stmaryrd}
\usepackage{algorithm2e}
\usepackage{url}
\usepackage{hyperref}  %affichage/création des liens hypertextes
\hypersetup{colorlinks,%
            citecolor=red,%
            filecolor=black,%
            linkcolor=blue,%
            urlcolor=blue}
\usepackage[backend=biber, style=ieee,doi=false,isbn=false,url=false,eprint=false,maxcitenames=2]{biblatex}
\usepackage[dvipsnames]{xcolor}
\usepackage{color}
\usepackage{soul}
\usepackage{cancel}

\newcommand{\E}[1]{\underset{#1}{\mathbb E}}

\newcommand{\modif}[1]{{\color{violet}#1}}

\newtheorem{remark}{Remark}

\begin{document}

\title{Automatic Tuning of Denoising Algorithms Parameters without Ground Truth}

\author{Arthur Floquet, Sayantan Dutta, \IEEEmembership{Member, IEEE},  Emmanuel Soubies, \\ Duong-Hung Pham, \IEEEmembership{Member, IEEE}, and Denis Kouame, \IEEEmembership{Senior Member, IEEE}

\vspace{-1mm}

\thanks{This work was supported by the CIMI Excellence Laboratory, ANR grant ANR-11-LABX-0040 within the French State
Programme “Investissements d’Avenir”}
\thanks{A. Floquet, E. Soubies, D.-H. Pham, and D. Kouam\'e are with the IRIT Laboratory,  Universit\'e de Toulouse, and CNRS, Toulouse 31400, France. Email:  firstname.lastname@irit.fr. (\textit{Corresponding author : Arthur Floquet})} 
\thanks{S. Dutta is with the Department of Radiology, Weill Cornell Medicine, New York, NY 10022, USA. Email:  sdu4004@med.cornell.edu.}
}

\markboth{}
{Shell \MakeLowercase{\textit{et al.}}: Bare Demo of IEEEtran.cls for IEEE Journals}
\maketitle

\begin{abstract}
Denoising is omnipresent in image processing. It is usually addressed with algorithms relying on a set of hyperparameters that control the quality of the recovered image. Manual tuning of those parameters can be a daunting task, which calls for the development of automatic tuning methods.
Given a denoising algorithm, the best set of parameters is the one that minimizes the error between denoised and ground-truth images. Clearly, this ideal approach is unrealistic, as the ground-truth images are unknown in practice. In this work, we propose unsupervised cost functions --- i.e., that only require the noisy image ---  that allow us to
reach this ideal gold standard performance.
Specifically, the proposed approach makes it possible to obtain an average PSNR output within less than 1\% of the best achievable PSNR.

\end{abstract}


\IEEEpeerreviewmaketitle

\vspace{-3mm}

\section{Introduction}
\IEEEPARstart{N}{oise} is inherent to any imaging device. It comes from a variety of sources and is modeled in a variety of ways. When considering additive noise, the corrupted measurements $\mathbf{y} \in \mathbb{R}^N$  follow the model : 
\begin{equation} 
\mathbf{y} = \mathbf{x} + \mathbf{n} ,
\label{intro:model}
\end{equation}
with $\mathbf{x} \in \mathbb{R}^N$ the clean image we wish to recover, and $\mathbf{n} \in \mathbb{R}^N$ the noise. Many denoising algorithms, denoted $A_{\bm{\uptheta}}$, have been proposed to address this task and provide estimates given by $A_{\bm{\uptheta}}(\mathbf{y})$, \textit{e.g},~\cite{dabov_image_nodate, selesnick_total_2017, dutta_novel_2022}, to cite few. The quality of these estimates depends on the chosen parameters $\bm{\uptheta} \in \mathbb{R}^d$.  In practice, however, manual tuning of these parameters is far from being trivial, even for low numbers of parameters such as two or three. As such, finding ways to automatically tune these parameters is of major importance and constitutes an active area of research.
Most existing approaches use a mapping $\bm{\Theta}_{\bm{\uplambda}}: \mathbf{y} \mapsto \bm{\uptheta}$, itself parameterized by $\bm{\uplambda} \in \mathbb{R}^{d'}$, that maps an image and/or its features (e.g., noise level, noise type, image dynamic, image content) to a set of parameters $\bm{\uptheta}$. The best $\bm{\uplambda}$, \textit{i.e.} $\bm{\uplambda}^*$, is found by minimizing the expectancy $\E{}$ of a discrepancy measure $\mathcal{L}(\cdot,\cdot)$ between the denoised images and ground truth images: 
\begin{equation}\label{eq:f_fit}
    {\bm{\uplambda}}^{*}:= \arg \min_ {\bm{\uplambda} \in \mathbb{R}^{d'}} \E{(\mathbf{x},\mathbf{y})} [\mathcal{L}(A_{\bm{\Theta}_{\bm{\uplambda}}(\mathbf{y})}(\mathbf{y}), \mathbf{x})] .
 \end{equation} 

There are several possibilities to define $\bm{\Theta}_{\bm{\uplambda}}$ with increasing degree of sophistication: 
\begin{itemize}
    \item $\bm{\Theta}_{\bm{\uplambda}}$ is a constant mapping (i.e., $\bm{\Theta}_{\bm{\uplambda}}(\mathbf{y}) = \bm{\uptheta} \in \mathbb{R}^d$). Solving~\eqref{eq:f_fit} then consists in finding a fixed set of parameters $\bm{\uptheta}$ such that, on average, the estimates are good, \textit{e.g},~\cite{calatroni_bilevel_2015}. 
    \item $\bm{\Theta}_{\bm{\uplambda}}$ is defined using features extracted from $\mathbf{y}$. For instance, $\bm{\Theta}_{\bm{\uplambda}}(\mathbf{y}) = \lambda_1 + \lambda_2 \hat{\sigma}(\mathbf{y})$ where $\hat{\sigma}(\mathbf{y})$ is an estimator of the noise variance. Here, solving~\eqref{eq:f_fit} allows us to learn a mapping $\bm{\Theta}_{\bm{\uplambda}}$ that adjusts algorithm parameters according to a \textit{known} and \textit{predefined} model, \textit{e.g},~\cite{nguyen_map-informed_nodate}. 
    \item $\bm{\Theta}_{\bm{\uplambda}}$ is a neural network. In this case, solving~\eqref{eq:f_fit} allows us to learn a mapping $\bm{\Theta}_{\bm{\uplambda}}$ that adjusts algorithm parameters according to an \textit{unknown} model, \textit{e.g},~\cite{afkham_learning_2021, kofler_learning_2023}. This greater flexibility comes at the price of a larger number of parameters $\bm{\uplambda}$ to learn.
\end{itemize}

Typically, all these methods require a dataset and work in a supervised way. Since this is not always feasible, unsupervised alternatives have been developed \cite{lehtinen_noise2noise_2018, xu_noisy-as-clean_2020, moran_noisier2noise_2020, pang_recorrupted--recorrupted_2021}. The main idea behind these approaches is to define an unsupervised loss (\textit{i.e.}, that does not depend on $\mathbf{x}$) achieving the same minimizers as the supervised counterpart. Nevertheless, constructing $\bm{\Theta}_{\bm{\uplambda}}$ and finding $\bm{\uplambda}$ remains challenging.

To discard the need of defining and training a mapping $\bm{\Theta}_{\bm{\uplambda}}$ on a dataset, one could  directly fit ${\bm{\uptheta}}$ on individual images~$\mathbf{y}$. The ideal estimate would be:
\begin{equation}
    \begin{split}
        & \mathbf{x}^{*}:= A_{\bm{\uptheta}^{*}}(\mathbf{y}) \\
        \text{with } & {\bm{\uptheta}}^{*}:= \arg \min_{\bm{\uptheta} \in \mathbb{R}^d} \mathcal{C}^*_{\mathbf{y}}(\bm{\uptheta}) ~,~ \mathcal{C}^*_{\mathbf{y}}(\bm{\uptheta}) := \mathcal{L}(A_{\bm{\uptheta}}(\mathbf{y}),\mathbf{x}).
    \end{split}
    \label{eq:gold_standard}
\end{equation}
Yet, this formulation is impractical as it requires knowing $\mathbf x$ to obtain ${\mathbf{x}}^*$. In the following, ${\mathbf{x}}^{*}$ will be our gold standard, that is, the best estimate we can expect for a given image $\mathbf{y}$ and algorithm $A_{\bm{\uptheta}}$. For Gaussian noise, methods such as  the famous  generalized cross-validation (GCV) \cite{Golub1979} and its variants or the Stein’s unbiased risk estimate (SURE)~\cite{stein1981estimation}, which depends only on the noisy data, can be used in place of the true mean-squared error (MSE). The SURE optimization is, however, challenging in the general case and requires the use of approximations~\cite{Deledalle_2014} or Monte Carlo approaches~\cite{ramani_monte-carlo_2008}. It is noteworthy to mention that other metrics that do not require the reference image have also been proposed~\cite{xiang_zhu_automatic_2010, SC_metric_2017}.

Here, inspired by~\cite{lehtinen_noise2noise_2018, xu_noisy-as-clean_2020, moran_noisier2noise_2020, pang_recorrupted--recorrupted_2021}, we propose alternative unsupervised cost functions $\widehat{\mathcal{C}}$ and inference schemes~$\widehat{\mathcal{I}}$  such that: 
\begin{equation}\label{eq:unsup_fit}
    \begin{split}
        & \mathbf{x}^{*} \approx \hat{\mathbf{x}}:= \widehat{\mathcal{I}}(\hat{\bm{\uptheta}}, \mathbf{y}) \\
        \text{with } & \hat{\bm{\uptheta}}:= \arg \min_{\bm{\uptheta} \in \mathbb{R}^d} \widehat{\mathcal{C}}_{\mathbf{y}}(\bm{\uptheta}).
    \end{split}
\end{equation}
Let us emphasize that the inference scheme in~\eqref{eq:unsup_fit} is not directly $A_{\hat{\bm{\uptheta}}}(\mathbf{y})$. Indeed, as detailed in Section~\ref{section:method}, the proposed unsupervised cost functions may require adapting the inference scheme. As such, we will systematically specify both the cost function and the inference scheme.

\begin{remark}
    The use of the cost function $\mathcal{C}^*$ could be avoided in \eqref{eq:gold_standard} and replaced with $\mathcal{L}$ directly. We write \eqref{eq:gold_standard} this way to be consistent with the following unsupervised cost functions $\widehat{\mathcal{C}}$, which don't reduce to simple discrepancy measures between two images.
Moreover, we will use $\cdot^*$ and $\widehat{\cdot}$ for supervised and unsupervised objects, respectively.
\end{remark}

\subsection*{Paper Outline}
The paper is structured as follows. The proposed method is detailed in Section \ref{section:method}. Then, in Section~\ref{section:results}, we deploy our approach to tune the parameters of a recently published denoiser~\cite{dutta_novel_2022} that can handle a variety of noise types. Finally, discussions and conclusions are provided in Section \ref{section:conclusion}.

\section{Method}
\label{section:method}
Let us start this section by restating the key difference between our approach in Eq.~\eqref{eq:unsup_fit} and the more standard one in Eq.~\eqref{eq:f_fit} (with its unsupervised counterparts~\cite{lehtinen_noise2noise_2018, xu_noisy-as-clean_2020, moran_noisier2noise_2020, pang_recorrupted--recorrupted_2021}). Notably, there is no training phase in~\eqref{eq:unsup_fit}. We don't need to create and train a mapping $\bm{\Theta}_{\bm{\uplambda}}$ over a dataset that will then be used to infer estimates on new data. Instead, our inference is done by solving (\ref{eq:unsup_fit}) directly for individual images $\mathbf{y}$.

We also emphasize that our aim is not to build a new denoiser, but an automatic way to select hyperparameters of a given denoiser, using only the input noisy image.  For illustration purposes, we use the DeQuIP algorithm~\cite{dutta_novel_2022} described in Section \ref{subsec:dequip}. 

\subsection{The Cost Functions and Inference Schemes Definition}
We drew our inspiration from existing unsupervised learning methods. Instead of training a denoising neural network $f$ of parameters $\upomega$, say $f_{\bm{\upomega}}$, on noisy-clean pairs $(\mathbf{y}, \mathbf{x})$, Noise2Noise (N2N) \cite{lehtinen_noise2noise_2018} proposes to train it on pairs of noisy images $(\mathbf{y}, \mathbf{y'})$ with $\mathbf{y}$ and $\mathbf{y'}$ two noisy versions of the same clean image $\mathbf{x}$. Lehtinen~\textit{et~al.}~\cite{lehtinen_noise2noise_2018} showed that for several noise types, $\mathcal{L}$ can be chosen accordingly so that:
\begin{equation}\label{eq:n2n_general}
   \mkern-8mu \arg \min_{\bm{\upomega}} \E{(\mathbf{y}, \mathbf{y'})}\mathcal{L}(f_{\bm{\upomega}}(\mathbf{y}), \mathbf{y'}) = \arg \min_{\bm{\upomega}} \E{(\mathbf{x}, \mathbf{y})} \mathcal{L}(f_{\bm{\upomega}}(\mathbf{y}), \mathbf{x}) .
\end{equation}
For example, if $\mathbf{y}, \mathbf{y'}$ are two corrupted versions of $\mathbf{x}$ with independent additive zero-mean noise (i.e., $\mathbf{y} = \mathbf{x} + \mathbf{n} $, $\mathbf{y'} = \mathbf{x} + \mathbf{n'}$ with $\mathbb E[\mathbf{n}] = \mathbb E[\mathbf{n'}] = \mathbf{0}$), letting $\mathcal{L}$ be the quadratic error leads to~\eqref{eq:n2n_general}: 
\begin{align}
 \nonumber
    \hat{{\bm{\upomega}}}  := & \arg \min_{\bm{\upomega}} \E{(\mathbf{y},\mathbf{\mathbf{y'}})} \|f_{\bm{\upomega}}(\mathbf{y}) - \mathbf{y'}\|^2_2 \\
    \nonumber
    = & \arg \min_{\bm{\upomega}} \E{(\mathbf{y},\mathbf{x},\mathbf{\mathbf{n}'})}  \|f_{\bm{\upomega}}(\mathbf{y}) - \mathbf{x} - \mathbf{n'}\|^2_2 \\
    \nonumber
    = & \arg \min_{\bm{\upomega}} \E{(\mathbf{y},\mathbf{x},\mathbf{\mathbf{n}'})} \left[ \|f_{\bm{\upomega}}(\mathbf{y}) - \mathbf{x}\|^2_2 - 2\left<f_{\bm{\upomega}}(\mathbf{y}) - \mathbf{x}, \mathbf{n}' \right>\right]\\
    = & \arg \min_{\bm{\upomega}} \E{(\mathbf{y},\mathbf{x})}  \|f_{\bm{\upomega}}(\mathbf{y}) - \mathbf{x}\|^2_2:= \bm{\upomega}^*,
\label{eq_proof}
\end{align}
where the expectation of the  dot product cancels as $\mathbb E[\mathbf{n'}] = \mathbf{0},$ and  $\mathbf{n}$ and $\mathbf{n'}$ (i.e., $\mathbf{y}$ and $\mathbf{n'}$) are independent.

In practice, training is performed using a finite dataset $\Omega_{I,J}:= \{ \mathbf{y}_i^j = \mathbf{x}_i + \mathbf{n}_i^j \}_{i,j = (1,1)}^{I,J}$. This means that for a  noiseless image $\mathbf{x}_i$, $J$ noisy versions are available, with additive noises $\mathbf{n}_i^j$ . As such, the true loss in~\eqref{eq:n2n_general} is replaced by an empirical one. This results in a solution $\hat{\bm{\upomega}} \approx \bm{\upomega}^{*}$ whose quality depends on $I,J$: 
\begin{itemize}
    \item When both $I$ and $J$ are large, the empirical loss yields a good approximation of the true one.
    \item When $I$ is small (e.g., $I = 1$), but $J$ is large, we can expect $f_{\widehat{\bm{\upomega}}}$ to perform well on the training set, but not to generalize well. It will effectively learn to output the mean of the $\mathbf{y}^j_1$, which is $\mathbf{x}_1$. 
    \item When $J$ is small (e.g., $J = 2$, with different $\mathbf{n}_i^1$, $\mathbf{n}_i^2$ for each $i$), $f_{\widehat{\bm{\upomega}}}$ could overfit the dataset $\Omega_{I,J}$ to the point that we systematically obtain $f_{\widehat{\bm{\upomega}}}(\mathbf{y}^1_i) = \mathbf{y}^2_i, \ f_{\widehat{\bm{\upomega}}}(\mathbf{y}^2_i) = \mathbf{y}^1_i, \ \forall i$. Yet, a large $I$ (e.g., $I = 1000$) seems to be sufficient in preventing overfitting as demonstrated by Lehtinen~\textit{et~al,}~\cite{lehtinen_noise2noise_2018}. Other factors can be exploited to avoid overfitting when limited data are available. These include network architecture, image type, noise type, or a low number of parameters relative to image size, $\mathrm{card}({\bm{\upomega}})/P$, with $P$ the number of image pixels~\cite{ying_overview_2019}. For instance, it has been shown that U-Net-like networks have a hard time recreating noise~\cite{ulyanov_deep_2020}.
    \item When both $I$ and $J$ are small (e.g., $I = 1$, $J = 2$), if not carefully designed, $f_{\bm{\upomega}}$ would simply learn to map $\mathbf{y}^1_1$ to $\mathbf{y}^2_1$, and vice-versa.
\end{itemize}
The proposed method, i.e., fitting algorithm parameters to a single image, corresponds to the last extreme case. We argue that if, instead of a neural network (which can recreate noise), we use a denoising algorithm $A_{\bm{\uptheta}}$ with low $\mathrm{card}({\bm{\uptheta}})/P$, we can define:
\begin{equation}\label{eq:n2n_loss}
         \widehat{\mathcal{C}}^{\mathrm{N2N}}_{\mathbf{y}, \mathbf{y'}}(\bm{\uptheta})  := \|A_{\bm{\uptheta}}(\mathbf{y}) - \mathbf{y'}\|^2_2 , \, 
         \hat{\mathbf{x}}^{\mathrm{N2N}} := A_{\hat{\bm{\uptheta}}}(\mathbf{y}),
\end{equation}
and have $\hat{\mathbf{x}}^{\mathrm{N2N}} \approx \mathbf{x}^{*}$ (we remind the reader that here, as for the rest of the paper, $\hat{\bm{\uptheta}}:= \arg \min \widehat{\mathcal{C}}_{\mathbf{y}}(\bm{\uptheta})$). This claim, consistent with (\ref{eq_proof}), is supported by our numerical experiments in Section~\ref{section:results} where we consistently obtain an estimate $\hat{\mathbf{x}}^{\mathrm{N2N}}$ of the same quality as the gold standard $\mathbf{x}^{*}$. 

\begin{remark}  
    The cost $\widehat{\mathcal{C}}^{\mathrm{N2N}}_{\mathbf{y}, \mathbf{y'}}$ is presented using the $L_2$ norm. Yet, other distances  $\mathcal{L}$  can be considered, for instance, to handle other types of noise as in~\cite{lehtinen_noise2noise_2018}.
\end{remark}

Still, N2N requires two independent noisy versions of the same clean image, which is not very standard in practice. Thus, we explored alternative ideas that extend the N2N one such that a single noise realization $\mathbf{y}$ per clean image $\mathbf{x}$ is required. They revolve around strategies of renoising $\mathbf{y}$ into $\mathbf{z}$ so as to create pairs of noisy images. These ideas and the corresponding costs and inferences we propose in our context are described below.

\paragraph*{Noisy-as-Clean (NaC) \cite{xu_noisy-as-clean_2020}} The underlying idea is to create a doubly noisy image $\mathbf{z} =\mathbf{y} + \mathbf{n}_\mathrm{s} $, with $\mathbf{n}_\mathrm{s}$ being simulated noise drawn from the same distribution as $\mathbf{n}$ according to (\ref{intro:model}), and learn to go from $\mathbf{z}$  to $\mathbf{y}$. Inference is then done on regular noisy images: $\mathbf{\hat{x}} = f_{\widehat{{\bm{\upomega}}}}(\mathbf{y})$. From the continuity of $\mathcal{L}$, $f_\omega$, $p_{\mathbf{y}|\mathbf x}$, $p_{\mathbf{z}|\mathbf{y}}$ and assuming that the noise is low, the authors in~\cite{xu_noisy-as-clean_2020} argue that $\widehat{{\bm{\upomega}}} \approx {\bm{\upomega}}^{*}$. Since $f_\omega$ is continuous, this implies $\hat{\mathbf{x}} \approx \mathbf{x}^*$. Building upon this idea, we propose the following scheme: 
    \begin{equation}\label{eq:nac_loss}
             \widehat{\mathcal{C}}^{\mathrm{NaC}}_{\mathbf{y}}(\bm{\uptheta}): =  \E{\mathbf{z}} \ \|A_{\bm{\uptheta}}(\mathbf{z}) - \mathbf{y}\|^2_2 ,\,
             \hat{\mathbf{x}}^{\mathrm{NaC}} := A_{\hat{\bm{\uptheta}}}(\mathbf{y}).
    \end{equation}
    
\paragraph*{Noisier2Noise (Nr2N) \cite{moran_noisier2noise_2020}} As in NaC, it also proceeds by creating a doubly noisy image $\mathbf{z} =\mathbf{y} + \mathbf{n}_\mathrm{s} $, with $\mathbf{n}_s$ following the same distribution as $\mathbf{n}$ according to (\ref{intro:model}), and learn to go from $\mathbf{z}$  to $\mathbf{y}$. However, it does not require any other assumptions than the sole additive noise. In particular, it is not restricted to a low noise level. As such, $\widehat{\bm{\upomega}} \not \approx \bm{\upomega}^{*}$, so inference cannot be performed on $\mathbf{y}$. It is instead done on $\mathbf{z}$ and includes a correction step: $\mathbf{\hat{x}} = 2f_{\widehat{{\bm{\upomega}}}}(\mathbf{z}) - \mathbf{z} =\mathbb E[\mathbf{x}|\mathbf{z}]$. This amounts to supervised denoising of the image $\mathbf{z}$, which has a higher noise level than $\mathbf{y}$. To mitigate the effect of this artificially high noise level, one solution is to lower the variance of $\mathbf{n}_s$. The correction step will then depend on the new variance. For Gaussian noise, with $\mathbf{n} \sim \mathcal{N}(0,\sigma)$, NaC uses $\mathbf{n}_s \sim \mathcal{N}(0, \alpha \sigma)$, with $\alpha \in ]0 , 1]$. The inference becomes $\mathbf{\hat{x}} = \frac{(1+ \alpha^2)f_{\widehat{{\bm{\upomega}}}}(\mathbf{z}) - \mathbf{z}}{\alpha^2}$. A tradeoff has then to be made between a lower simulated noise, and an amplification of denoising errors due to the $\alpha^2$ division. Exploiting the Nr2N idea in our context leads to %the following scheme: 
    \begin{equation}\label{eq:nr2n_loss}
    \begin{split}
        & \widehat{\mathcal{C}}^{\mathrm{Nr2N}}_{\mathbf{y}}(\bm{\uptheta}) :=  \E{\mathbf{z}} \ \|A_{\bm{\uptheta}}(\mathbf{z}) - \mathbf{y}\|^2_2 \\
        & \hat{\mathbf{x}}^{\mathrm{Nr2N}} := \frac{(1 + \alpha^2)A_{\hat{\bm{\uptheta}}}(\mathbf{z}) - \mathbf{z}}{\alpha^2}.
    \end{split}
    \end{equation}
    
\paragraph*{Recorrupted-to-Recorrupted (R2R) \cite{pang_recorrupted--recorrupted_2021}} Here, the authors propose to create two doubly noisy images $\mathbf{z}_1 = \mathbf{y} + \mathbf{D}^\mathrm{T} \mathbf{n}_\mathrm{s} $ and $\mathbf{z}_2 = \mathbf{y} - \mathbf{D^{-1}}\mathbf{n}_\mathrm{s} $, with $\mathbf D$ being any invertible matrix and $\mathbf{n}_\mathrm{s}$ drawn from the same distribution as $\mathbf{n}$, according to~\eqref{intro:model}. Note that for R2R, $\mathbf{n}$ is assumed to be Gaussian. This way, we can write $\mathbf{z}_1 = \mathbf{x} + \mathbf{n}_1$, $\mathbf{z}_2 = \mathbf{x} + \mathbf{n}_2$, with $\mathbf{n}_1$ and $\mathbf{n}_2$ being zero-mean, independent noise vectors. Training can then be done as in N2N. Yet, as in NaC and Nr2N, such unsupervised training can be seen as supervised training with a higher noise level. To deal with this, R2R uses a Monte-Carlo scheme in inference to limit the effect of $\mathbf{n}_s $: $\mathbf{\hat{x}} = \frac{1}{M} \sum_m f_{\widehat{\bm{\upomega}}}(\mathbf{z}^m_1)$ where the superscript $m$ simply represents the $m^{\mathrm{th}}$ drawing of $\mathbf{n}_s$. For our fitting problem~\eqref{eq:unsup_fit}, we propose the  scheme: 
    \begin{equation}\label{eq:r2r_loss}
        \begin{split}
            & \widehat{\mathcal{C}}^{\mathrm{R2R}}_{\mathbf{y}}(\bm{\uptheta}):= \E{(\mathbf{z}_1, \mathbf{z}_2)}\|A_{\bm{\uptheta}}(\mathbf{z}_1) - \mathbf{z}_2\|^2_2 \\
            & \hat{\mathbf{x}}^{\mathrm{R2R}} := \frac{1}{M} \sum_{m=1}^M A_{\hat{\bm{\uptheta}}}(\mathbf{z}^m_1).
        \end{split}
    \end{equation}

\subsection{Optimization Aspects}
To solve Problem~\eqref{eq:unsup_fit}, we exploit automatic differentiation to perform gradient descent steps.  This implies that $A_{\bm{\uptheta}}$ must be differentiable with respect to $\bm{\uptheta}$, or at least subdifferentiable. For instance, as explained in Section \ref{subsec:dequip}, hard thresholds need to be replaced with soft thresholds. For $\widehat{\mathcal{C}}^{\mathrm{NaC}}$, $\widehat{\mathcal{C}}^{\mathrm{Nr2N}}$ and $\widehat{\mathcal{C}}^{\mathrm{R2R}}$, we approximate $\mathbb{E}_{\mathbf z}$ by drawing a new random $\mathbf{z}^k$ at each iteration $k \in \llbracket 1 ; K \rrbracket$.
For initialization, we use a fixed $\bm{\uptheta}_0$, found by manually tuning $\bm{\uptheta}$ for \textit{a single image}. \footnote{\modif{\href{https://gitlab.com/a_floquet/automatic-tuning-of-denoising-algorithms-parameters-without-ground-truth}{Codes and supplementary materials can be found here}}}

\noindent 

%%%%%%%%%%%%%%%%% This figure is defined here to go on top of the next page %%%%%%%%%%%%%%%%%%%%%%%%%%%%
\begin{figure*}[t]
    \centering
    \includegraphics{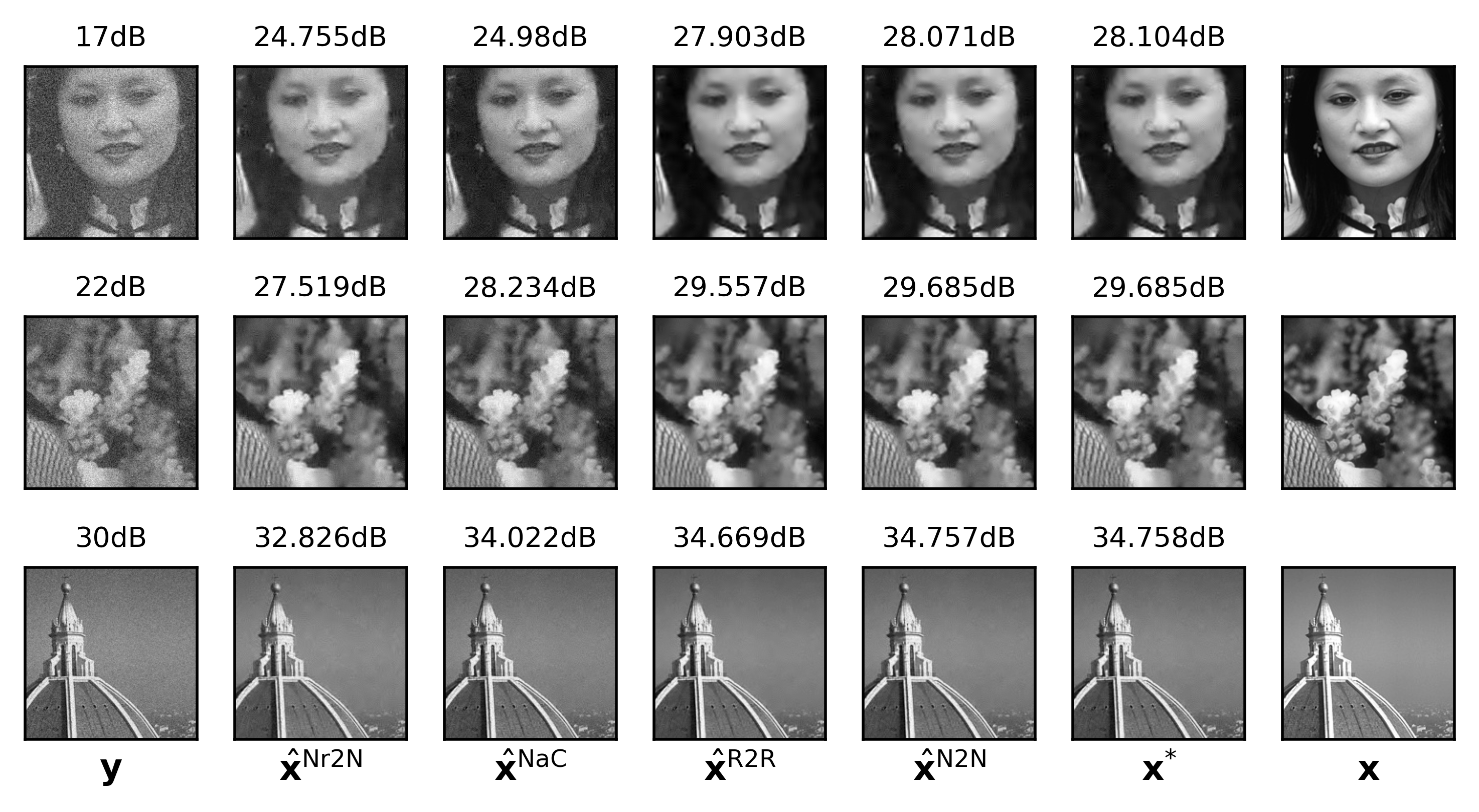}
    \vspace{-0.2cm}
    \caption{Exemple of denoising results. The upper scripts of $\mathbf{\hat{x}}$ refer to the corresponding method; $\mathbf{x}^*$ and $\mathbf{x}$ denote respectively the result for the supervised (gold standard) approach and the ground-truth.}
    \label{fig:comp_im}
\end{figure*}
%%%%%%%%%%%%%%%%%%%%%%%%%%%%%%%%%%%%%%%%%%%%%%%%%%%%%%%%%%%%%%%%%%%%%%%%%%%%%%%%%%%%%%%%%%%%%%%%%%%%%%%%%

\subsection{Use case: Denoising via Quantum Interactive Patches}\label{subsec:dequip} %
We illustrate our method, without loss of generality, to a denoiser, denoted DeQuIP (\textit{Denoising via Quantum Interactive Patches})~\cite{dutta_novel_2022} which can handle various types of noise.
It is based on two main concepts of quantum mechanics: (i) exploits the Schr\"odinger equation of quantum physics to construct an adaptive basis, which enables DeQuIP to deal with various noise models; (ii) treats the image as patches, and formalizes the self-similarity between neighbor patches through a term akin to quantum many-body interaction to efficiently preserve the local structures. Each patch behaves as a single-particle system, which interacts with other neighboring patches, thus the whole image acts as a many-body system.
Under a potential ${\mathbf{V}}$,  represented by the image pixel values \cite{dutta_quantum_2021}, the adaptive basis vector ${\bm{\uppsi}}$ describes the characteristics of a virtual quantum particle with energy $E$, and satisfies the non-relativistic stationary Schr\"odinger equation, written as: 
\begin{equation}
    - \frac{\hbar ^2}{2m} \nabla ^2 {\bm{\uppsi}} = - {\mathbf{V}} {\bm{\uppsi}} + E {\bm{\uppsi}},
\label{eq:schro}
\end{equation}
with $\hbar$ being Planck's constant, $m$ the particle mass, and $\nabla ^2$ the Laplacian operator. The image-dependent basis is constructed from the wave solutions ${\bm{\uppsi_i}}$ of eq.~\eqref{eq:schro} by plugging the noisy image $\mathbf{y}$ as the potential ${\mathbf{V}}$ of the system. These wave solutions are oscillating functions with a local frequency proportional to $\sqrt{(E - {\mathbf{V}})/(\hbar ^2/2m)}$, thus the frequency is locally adapted to the image pixels' values.
The exact behavior of these basis vectors with respect to the potential is determined by the constant $\frac{\hbar^2}{2m}$, which is a parameter of DeQuIP.
For a more detailed illustration of these oscillating wave vectors, we refer readers to \cite{dutta_quantum_2021, dutta_image_2021}. Finally, for denoising, the noisy image $\mathbf{y}$ is projected onto this adaptive basis, and the low-value coefficients are thresholded using a soft threshold: 
\begin{equation*}
    \Upsilon_{c_1, c_2}(x) = \left\{
    \begin{matrix*}[l]
        0 & \text{if } |\alpha |\leq c_1 \\
        \frac{c_2}{c_2 - c_1}  x - \frac{c_1 c_2}{c_2 - c_1}  & \text{if } -c_2 \leq x \leq -c_1 \\
        \frac{c_2}{c_2 - c_1}  x + \frac{c_1 c_2}{c_2 - c_1}  & \text{if }  c_2 \geq x \geq c_1 \\
        1 & \text{elsewhere}
    \end{matrix*}\right.
\end{equation*}

The second key idea of DeQuIP is the integration of quantum interaction theory to consider the image as overlapping patches, where each patch is a single-particle system that interacts with other patches \cite{dutta_novel_2022}. The interaction between two patches $\mathbf{A}$ and $\mathbf{B}$ is defined as: ${\mathbf{I}}_{AB} = p \frac{|{\mathbf{A}} - {\mathbf{B}}|}{d(A,B)^2}$ with $p$ a parameter controlling the strength of the interaction term, and $d$ the euclidean pixel distances between the center of $\mathbf{A}$ and $\mathbf{B}$. This term is designed to promote local self-similarity, based on the hypothesis that neighboring patches in an image are likely to be more similar than distant ones.

Therefore, DeQuIP has $4$ parameters $\bm{\uptheta} = \{ -\frac{\hbar^2}{2m}, p, c_1, c_2 \}$  shared across patches.

\section{Experimental results}\label{section:results}

\subsection{Material \& Data}
Implementation is done on \textit{PyTorch} 1.12.0. We used $180 \times 180$ clean grayscale photographic images, extracted from the \textit{BSD400} dataset, as ground truth.

\subsection{Comparison of Denoised Images}\label{subsec:comparison_losses}
\begin{figure}[t]
    \centering
    \includegraphics[width = 0.47\textwidth]{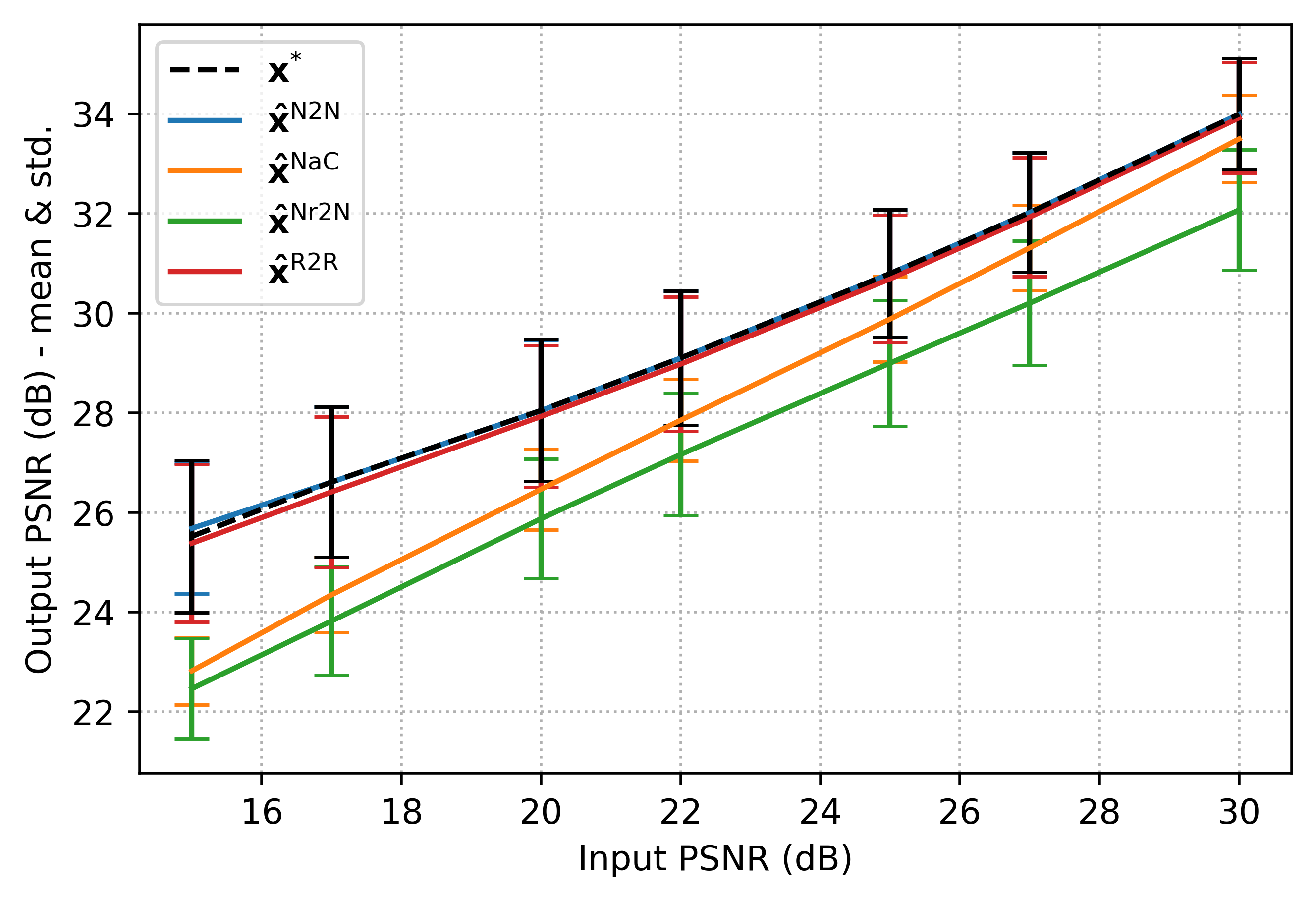}
    \vspace{-0.2cm}
    \caption{Comparison of the proposed strategies with the gold standard $\mathbf{x}^*$. The average performance and standard deviation is computed over 35 test images.}
    \label{fig:comp_plot}
\end{figure}

The estimates $\hat{\mathbf{x}}^{\mathrm{N2N}}$, $\hat{\mathbf{x}}^{\mathrm{NaC}}$, $\hat{\mathbf{x}}^{\mathrm{Nr2N}}$, $\hat{\mathbf{x}}^{\mathrm{R2R}}$ and the gold standard $\mathbf{x}^{*}$ are compared in the zero-mean Gaussian noise case since it is the only case covered by the four proposed schemes. 
We report on Fig.~\ref{fig:comp_plot} the average performance (measured by output PSNR) and standard deviation over 35 test images as a function of the input PSNR. Examples of noisy and denoised images are presented in Fig.~\ref{fig:comp_im}.
We see, both in metric and visually, that $\hat{\mathbf{x}}^{\mathrm{N2N}}$ and $\hat{\mathbf{x}}^{\mathrm{R2R}}$ achieve the gold standard performance $\mathbf{x}^*$. As expected given the assumptions it relies on, $\hat{\mathbf{x}}^{\mathrm{NaC}}$ yields better results when noise is low enough (i.e., high PSNR), but still underperforms. Finally, $\hat{\mathbf{x}}^{\mathrm{Nr2N}}$ does not perform as well as the other unsupervised methods. 

\subsection{Poisson Noise Denoising}
\begin{figure}[t]
    \centering
    \includegraphics[width = 0.495\textwidth]{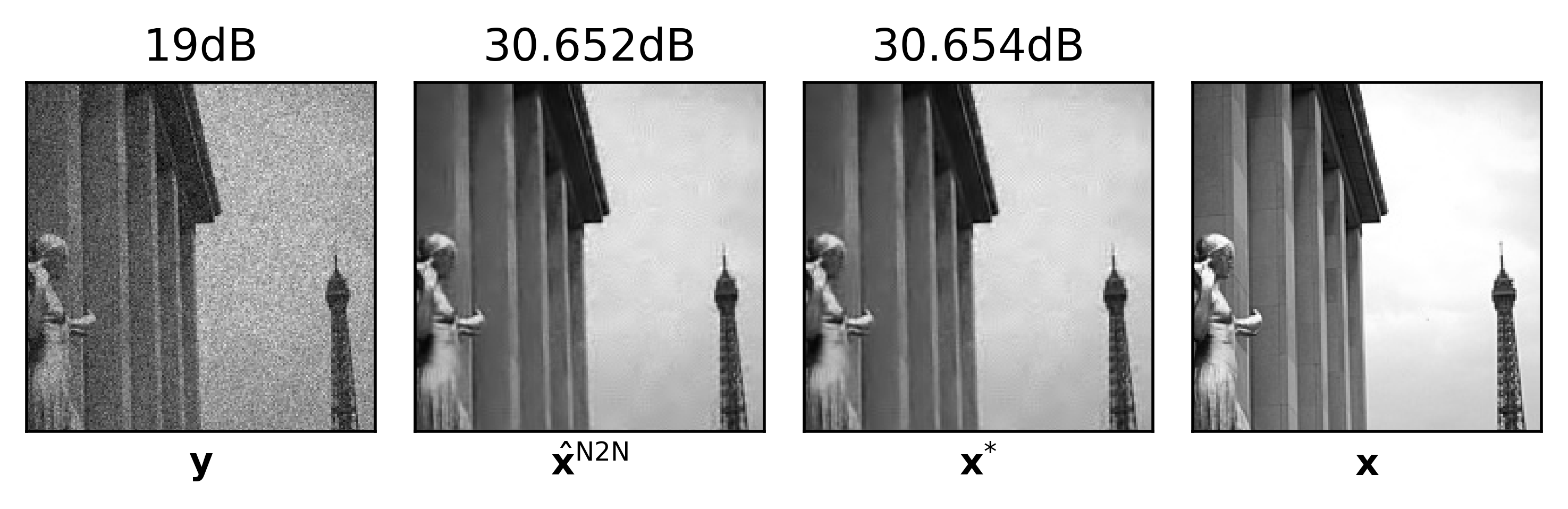}
    \caption{Poisson noise denoising.}
    \label{fig:poisson}
\end{figure}

As DeQuIP can deal with various noise models, we present in Fig.~\ref{fig:poisson} Poisson noise denoising results, where we can observe that $\hat{\mathbf{x}}^{\mathrm{N2N}}$ is similar to $\mathbf{x}^{*}$. We used $\widehat{\mathcal{C}}^{\mathrm{N2N}}$ because of its performances, and its capacity to handle Poisson noise.

\section{Conclusion}\label{section:conclusion}
We have proposed a method for the automatic tuning of denoising algorithm parameters, leveraging only the noisy measurements targeted for enhancement. Specially, we introduced several cost functions and inference schemes, two of which yielded results comparable to those obtained with ground truth-tuned parameters. \\
However, our method comes with certain limitations. The first one arises when only a single noisy image is available, requiring the noise to follow a zero-mean Gaussian distribution. For other types of noise, two independent noisy versions are needed. The second limitation is the need to resolve an optimization problem that involves differentiating the denoising algorithm with respect to its parameters. Although this can be achieved using a simple backpropagation, it can be computationally expensive. \\
Moving forward, there are potential avenues for exploration. For algorithms $A$ that cannot replicate the identity function, exploring unsupervised losses inspired by Noise2Self  \cite{krull_noise2void_2019} and \cite{batson_noise2self_2019} could be fruitful, given their advantage of using a single noise realization $\mathbf{y}$ without renoising it. Another idea worth exploring is the extension of our method to other inverse problems such as image deblurring or super-resolution, possibly building on \cite{degraded2degraded, multi_operator_imaging}.

\printbibliography

@inproceedings{lehtinen_noise2noise_2018,
  url = {https://arxiv.org/abs/1803.04189},
  author = {Lehtinen,  Jaakko and Munkberg,  Jacob and Hasselgren,  Jon and Laine,  Samuli and Karras,  Tero and Aittala,  Miika and Aila,  Timo},
  title = {{Noise2Noise}: Learning Image Restoration without Clean Data},
  publisher = {PMLR},
  booktitle = { International Conference on Machine Learning},
  year = {Mar. 2018},
  doi = {10.48550/arXiv.1803.04189}
}

@article{xu_noisy-as-clean_2020,
    author={Xu, Jun and Huang, Yuan and Cheng, Ming-Ming and Liu, Li and Zhu, Fan and Xu, Zhou and Shao, Ling},
    journal={IEEE Transactions on Image Processing}, 
    title={{Noisy-as-Clean}: Learning Self-Supervised Denoising From Corrupted Image}, 
    year={2020},
    volume={29},
    number={},
    pages={9316-9329},
    doi={10.1109/TIP.2020.3026622}
}

@inproceedings{moran_noisier2noise_2020,
    doi = {10.48550/arxiv.1910.11908},
    author = {Moran,  Nick and Schmidt,  Dan and Zhong,  Yu and Coady,  Patrick},
    title = {{Noisier2Noise}: Learning to Denoise from Unpaired Noisy Data},
    publisher = {IEEE/CVF},
    booktitle = {Conference on Computer Vision and Pattern Recognition}, 
    year = {Oct. 2019}
}

@inproceedings{pang_recorrupted--recorrupted_2021,
    author={Pang, Tongyao and Zheng, Huan and Quan, Yuhui and Ji, Hui},
    booktitle={Conference on Computer Vision and Pattern Recognition}, 
    title={{Recorrupted-to-Recorrupted}: Unsupervised Deep Learning for Image Denoising}, 
    year={2021},
    volume={},
    number={},
    publisher={IEEE/CVF},
    pages={2043-2052},
    doi={10.1109/CVPR46437.2021.00208}
}

@article{afkham_learning_2021,
    title={Learning regularization parameters of inverse problems via deep neural networks},
    author={Afkham, Babak Maboudi and Chung, Julianne and Chung, Matthias},
    journal={Inverse Problems},
    volume={37},
    number={10},
    pages={105017},
    year={2021},
    publisher={IOP Publishing}, 
    doi={10.1088/1361-6420/ac245d}
}

@article{calatroni_bilevel_2015,
  title={Bilevel approaches for learning of variational imaging models},
  author={Calatroni, Luca and Cao, Chung and De Los Reyes, Juan Carlos and Sch{\"o}nlieb, Carola-Bibiane and Valkonen, Tuomo},
  journal={Variational Methods: In Imaging and Geometric Control},
  volume={18},
  number={252},
  pages={2},
  year={2017},
  publisher={Walter de Gruyter GmbH},
  doi={10.1515/9783110430394-008}
}

@article{dutta_novel_2022,
    doi = {10.1016/j.sigpro.2022.108690},
    url = {https://doi.org/10.1016/j.sigpro.2022.108690},
    year = {2022},
    month = dec,
    publisher = {Elsevier {BV}},
    volume = {201},
    pages = {108690},
    author = {Sayantan Dutta and Adrian Basarab and Bertrand Georgeot and Denis Kouam{\'{e}}},
    title = {A Novel Image Denoising Algorithm Using Concepts of Quantum Many-Body Theory},
    journal = {Signal Processing}
}

@article{kofler_learning_2023,
    title={Learning Regularization Parameter-Maps for Variational Image Reconstruction using Deep Neural Networks and Algorithm Unrolling},
    author={Kofler, Andreas and Altekr{\"u}ger, Fabian and Ba, Fatima Antarou and Kolbitsch, Christoph and Papoutsellis, Evangelos and Schote, David and Sirotenko, Clemens and Zimmermann, Felix Frederik and Papafitsoros, Kostas},
    journal={arXiv Preprint},
    year={Jan. 2023}, 
    DOI={10.48550/arXiv.2104.06594}
}

@article{krull_noise2void_2019,
    title={Noise2void-learning denoising from single noisy images},
    author={Krull, Alexander and Buchholz, Tim-Oliver and Jug, Florian},
    journal={Conference on Computer Vision and Pattern Recognition},
    publisher={IEEE/CVF},
    pages={2129--2137},
    year={2019}, 
    doi = {10.48550/arXiv.1811.10980}
}

@article{batson_noise2self_2019,
    title={Noise2self: Blind denoising by self-supervision},
    author={Batson, Joshua and Royer, Loic},
    journal={International Conference on Machine Learning},
    pages={524--533},
    year={2019},
    organization={PMLR}, 
    doi = {0.48550/arXiv.1901.11365}
}

@article{stein1981estimation,
  title={Estimation of the mean of a multivariate normal distribution},
  author={Stein, Charles M},
  journal={The annals of Statistics},
  pages={1135--1151},
  year={1981},
  publisher={JSTOR}
}

@inproceedings{ulyanov_deep_2020,
  doi = {10.1007/s11263-020-01303-4},
  year = {2018},
  publisher = {IEEE/CVF},
  pages = {1867--1888},
  author = {Dmitry Ulyanov and Andrea Vedaldi and Victor Lempitsky},
  title = {{Deep Image Prior}},
  booktitle = {Conference on Computer Vision and Pattern Recognition} 
}

@article{dabov_image_nodate,
    author={Dabov, Kostadin and Foi, Alessandro and Katkovnik, Vladimir and Egiazarian, Karen},
    journal={IEEE Transactions on Image Processing}, 
    title={Image Denoising by Sparse {3-D} Transform-Domain Collaborative Filtering}, 
    year={2007},
    volume={16},
    number={8},
    pages={2080-2095},
    doi={10.1109/TIP.2007.901238}
    }

@inproceedings{nguyen_map-informed_nodate,
title = {{MAP}-informed Unrolled Algorithms for Hyper-parameter Estimation},
    year={2023},
    month = {9},
    publisher = {IEEE},
    booktitle = {International Conference on Image Processing},
    author = {Nguyen, Pascal and Soubies, Emmanuel and Chaux, Caroline}
}

@article{selesnick_total_2017,
    author={Selesnick, Ivan},
    journal={IEEE Signal Processing Letters}, 
    title={Total Variation Denoising Via the Moreau Envelope}, 
    year={2017},
    volume={24},
    number={2},
    pages={216-220},
    doi={10.1109/LSP.2017.2647948}}

@ARTICLE{dutta_quantum_2021,
  author={Dutta, S. and Basarab, A. and Georgeot, B. and Kouamé, D.},
  journal={IEEE Open Journal of Signal Processing}, 
  title={Quantum Mechanics-Based Signal and Image Representation: Application to Denoising}, 
  year={2021},
  volume={2},
  number={},
  pages={190-206},
  doi={10.1109/OJSP.2021.3067507}
}

@INPROCEEDINGS{dutta_image_2021,
  author={Dutta, S. and Basarab, A. and Georgeot, B. and Kouamé, D.},
  booktitle={2021 IEEE International Conference on Image Processing (ICIP)}, 
  title={Image Denoising Inspired by Quantum Many-Body physics}, 
  year={2021},
  volume={},
  number={},
  pages={1619-1623},
  doi={10.1109/ICIP42928.2021.9506794}
}

@ARTICLE{Deledalle_2014,
  author={Deledalle, C. and Vaiter, S. and Fadili, J.and and Peyré, G.},
  journal={SIAM Journal on Imaging Sciences}, 
  title={Stein Unbiased GrAdient estimator of the Risk (SUGAR) for Multiple Parameter Selection}, 
  year={2014},
  volume={7},
  number={2},
  pages={2448-2487},
  doi={}
}

@article{ying_overview_2019,
	title = {An {Overview} of {Overfitting} and its {Solutions}},
	volume = {1168},
	issn = {1742-6588, 1742-6596},
	url = {https://iopscience.iop.org/article/10.1088/1742-6596/1168/2/022022},
	doi = {10.1088/1742-6596/1168/2/022022},
	language = {en},
	urldate = {2023-08-30},
	journal = {Journal of Physics: Conference Series},
	author = {Ying, Xue},
	month = feb,
	year = {2019},
	pages = {022022},
}

@article{ramani_monte-carlo_2008,
	title = {Monte-{Carlo} {Sure}: {A} {Black}-{Box} {Optimization} of {Regularization} {Parameters} for {General} {Denoising} {Algorithms}},
	volume = {17},
	issn = {1057-7149},
	shorttitle = {Monte-{Carlo} {Sure}},
	url = {http://ieeexplore.ieee.org/document/4598837/},
	doi = {10.1109/TIP.2008.2001404},
	language = {en},
	number = {9},
	urldate = {2023-09-22},
	journal = {IEEE Transactions on Image Processing},
	author = {Ramani, S. and Blu, T. and Unser, M.},
	month = sep,
	year = {2008},
	pages = {1540--1554},
}

@article{xiang_zhu_automatic_2010,
	title = {Automatic {Parameter} {Selection} for {Denoising} {Algorithms} {Using} a {No}-{Reference} {Measure} of {Image} {Content}},
	volume = {19},
	issn = {1057-7149, 1941-0042},
	url = {http://ieeexplore.ieee.org/document/5484579/},
	doi = {10.1109/TIP.2010.2052820},
	language = {en},
	number = {12},
	urldate = {2023-09-22},
	journal = {IEEE Transactions on Image Processing},
	author = {{Xiang Zhu} and Milanfar, Peyman},
	month = dec,
	year = {2010},
	pages = {3116--3132},
}

@article{Golub1979,
	title = {Generalized cross-validation
as a method for choosing a good ridge parameter,},
	volume = {21},
	issn = {1057-7149, 1941-0042},
	url = {http://ieeexplore.ieee.org/document/5484579/},
	doi = {10.1109/TIP.2010.2052820},
	language = {en},
	number = {12},
	urldate = {2023-09-22},
	journal = {Technometrics,},
	author = {{G. H, Golub,} and M, Heath and G, Wahba},
	month = May,
	year = {1979},
	
}

@inproceedings{degraded2degraded,
 author = {Xia, Zhihao and Chakrabarti, Ayan},
 booktitle = {Advances in Neural Information Processing Systems},
 editor = {H. Wallach and H. Larochelle and A. Beygelzimer and F. d\textquotesingle Alch\'{e}-Buc and E. Fox and R. Garnett},
 pages = {},
 publisher = {Curran Associates, Inc.},
 title = {Training Image Estimators without Image Ground Truth},
 url = {https://proceedings.neurips.cc/paper_files/paper/2019/file/0ed9422357395a0d4879191c66f4faa2-Paper.pdf},
 volume = {32},
 year = {2019}
}

@inproceedings{multi_operator_imaging,
title={Unsupervised Learning From Incomplete Measurements for Inverse Problems},
author={Tachella, Juli{\'a}n and Chen, Dongdong and Davies, Mike},
booktitle={Proceedings of the 36th Conference on Neural Information Processing Systems},
year={2022}}

@Article{SC_metric_2017,
author={Kong, Xiangfei
and Yang, Qingxiong},
title={No-Reference Image Quality Assessment for Image Auto-Denoising},
journal={International Journal of Computer Vision},
year={2018},
month={May},
day={01},
volume={126},
number={5},
pages={537-549},
issn={1573-1405},
doi={10.1007/s11263-017-1054-2},
url={https://doi.org/10.1007/s11263-017-1054-2}
}

\end{document}

\maketitle \vspace{1.5cm}

%******* Intro *****************
\large{\noindent This is a supplementary material file for the article: Automatic Tuning of Denoising Algorithms Parameters without Ground Truth. \\

\noindent GitLab repository of this article can be found \href{https://gitlab.com/a_floquet/automatic-tuning-of-denoising-algorithms-parameters-without-ground-truth}{here}.}

%******* Algorithms ************
\section{Detailled algorithms}
\begin{figure*}[h!]
    \centering

    \begin{forest}
for tree = {% settings for elemts in tree
% nodes
    draw, rounded corners,
    top color=white, bottom color=blue!20,
    font = \ttfamily,
    minimum height=2ex, text depth = 0.25ex,
   anchor = north, 
% edges 
     edge = {-Stealth},
    s sep = 1.5em,
    l sep = 6em
            },
EL/.style = {% shortenes for Edge Label, defined as style
   before typesetting nodes={% edge labels positioning
where n=1{edge label/.wrap value={node[pos=0.5,anchor=east]{#1}}}% above left
         {edge label/.wrap value={node[pos=0.5,anchor=west]{#1}}}% above right
                            }
            }% end of EL
% tree body
[At least 2 independant noisy version of image available ?, root % rot is style defined in document preamble
    [\large{${\hat{\mathbf{x}}}^\mathrm{N2N}$, algorithm \ref{alg:n2n}}, EL=Yes]
    [Zero mean Gaussian Noise ?, EL=No, root
        [\large{${\hat{\mathbf{x}}}^\mathrm{R2R}$, algorithm \ref{alg:r2r}}, EL=Yes]
        [Weak noise ?, EL=No, root
            [\large{${\hat{\mathbf{x}}}^\mathrm{NaC}$ algorithm \ref{alg:nac}}, EL=Yes]
            [Additive noise ?, EL=No, root
                [\large{${\hat{\mathbf{x}}}^\mathrm{Nr2N}$, algorithms \ref{alg:nr2n} \& \ref{alg:nr2n_gaussian}}, EL=Yes]
                [\large{Not traited in this work}, EL=No]
            ]
        ]
    ]
]
    \end{forest}
    
    \caption{Help in choosing appropriate method}
    \label{fig:tree}
\end{figure*}
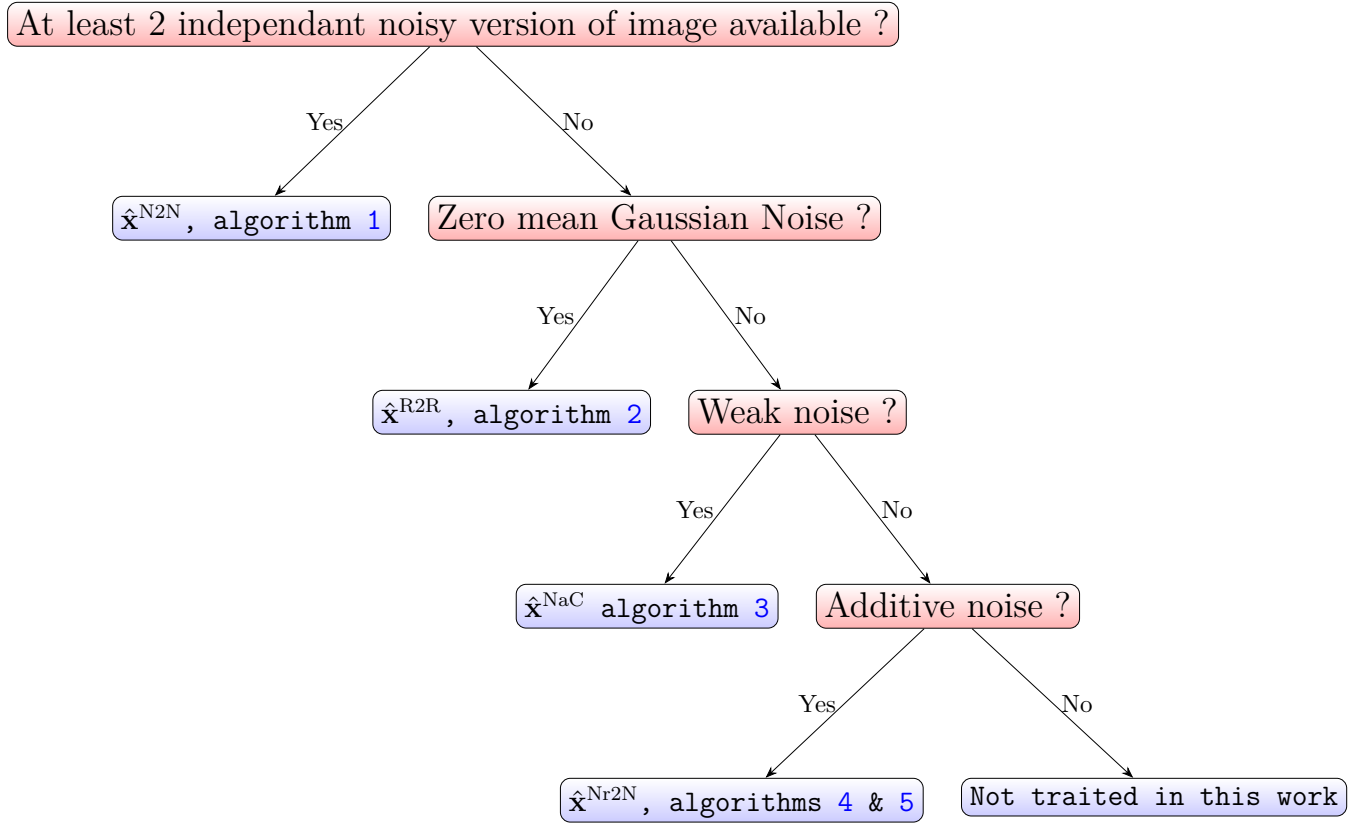

In the article, we presented four methods to tune the algorithm parameters $\bm{\uptheta}$. 
Here, we present a decision tree (Fig.~\ref{fig:tree}) allowing one to choose the most suitable method according to the available data, as well as the considered denoising algorithm. \\
Note that Fig.~\ref{fig:tree} includes all methods presented in the article. 
Empirically, we found ${\hat{\mathbf{x}}}^\mathrm{Nr2N}$ to perform poorly, and ${\hat{\mathbf{x}}}^\mathrm{NaC}$ to be acceptable only 
for small noise levels.
We believe ${\hat{\mathbf{x}}}^\mathrm{NaC}$ might be a suitable candidate for denoisers that somehow don't have parameters related to noise level.

\section{Hyperparameters}
The proposed method relies on some hyperparameters that include the learning rate, the number of iterations, the initial point $\bm{\uptheta}_0$. For the results reported in Fig.~2 of the main article, we fixed the number of iterations to $100$ and considered Adam optimizer with default parameters and a learning rate of $1$.  
Moreover, for $\hat{\mathbf{x}}^{\mathrm{Nr2N}}$, we fixed $\alpha = 1$, and for $\hat{\mathbf{x}}^{\mathrm{R2R}}$ we used $\mathbf D = \frac{1}{2} \mathbf I$, and $M=50$. 
Regarding the initial point $\bm{\uptheta}_0$ we fixed it through coarse manual tuning on a \textit{single image}. Yet, we were able to obtain high-quality denoised images which show the robustness of the approach to this initialization. Finally,  for $\mathbf{D}$ and $M$ in R2R, we followed the recommendation from \cite{pang_recorrupted--recorrupted_2021}. For Nr2N, the hyperparameter $\alpha$ was chosen equal to $1$ for simplicity. 

\textit{The take away message here is that (at least with the tested DeQuip denoising method) results were not sensitive to the choice of these parameters, which makes them easy to set.}

%%% Decision tree %%%%%

%%% Algo N2N %%%%%
\begin{algorithm}[h!]
\caption{Computing ${\hat{\mathbf{x}}}^\mathrm{N2N}$}\label{alg:n2n}
\begin{algorithmic}
\Require \\
Denoising algorithm $A_{\bm{\uptheta}}$ \\
$\mathbf{y}_1 ,\mathbf{y}_2$ two noisy version of $\mathbf{x}$,  with independant noises. \\
Initial parameters $\bm{\uptheta}_0$ \\
Learning rate $l_r$ \\
Discrepancy  measure $\mathcal L$ \Comment{Adapted to noise type~\cite{lehtinen_noise2noise_2018}} \\
Optimizer $\mathcal{O}$ \\
Number of iterations $K$ 
\For{$k \gets 1$ to $K$ }
    \State $\mathbf{u}_k \gets A_{\bm{\uptheta}_k}(\mathbf{y}_1)$
    \State Compute $\mathcal{L}(\mathbf{u}_k, \mathbf{y}_2)$
    \State Compute $\frac{\partial \mathcal{L}}{\partial \bm{\uptheta}_k}$
    \State $\bm{\uptheta}_{k+1} \gets \mathcal{O}\left( \bm{\uptheta}_{k}, l_r, \frac{\partial \mathcal{L}}{\partial \bm{\uptheta}_k} \right)$ \Comment{Optimizer step}
\EndFor
\State ${\hat{\mathbf{x}}}^\mathrm{N2N} \gets A_{\bm{\uptheta}_K}(\mathbf{y}_1)$
\end{algorithmic}
\end{algorithm}
%
%
%%% Algo R2R %%%%%
\begin{algorithm}[t]
\caption{Computing ${\hat{\mathbf{x}}}^\mathrm{R2R}$}\label{alg:r2r}
\begin{algorithmic}
\Require \\
Denoising algorithm $A_{\bm{\uptheta}}$ \\
$\mathbf{y} = \mathbf{x} + \mathbf{n}$, $\mathbf{n} \sim \mathcal{N}(\mathbf{0},\sigma \mathbf{I})$ \\
Invertible matrix $\mathbf{D}$ \\
Initial parameters $\bm{\uptheta}_0$ \\
Learning rate $l_r$ \\
Optimizer $\mathcal{O}$ \\
Number of iterations $K$ \\ 
Inference number $M$ 

\For{$k \gets 1$ to $K$ }
    \State $\mathbf{n}_s^k \gets \mathcal{N}(\mathbf{0}, \sigma \mathbf{I})$
    \State $\mathbf{z}_1^k = \mathbf{y} + \mathbf{D}^T \mathbf{n}_s^k$, $\mathbf{z}_2^k = \mathbf{y} - \mathbf{D}^{-1} \mathbf{n}_s^k$
    \State $\mathbf{u}_k \gets A_{\bm{\uptheta}_k}(\mathbf{z}_1^k)$
    \State $\mathcal{L} \gets ||\mathbf{u}_k - \mathbf{z}_2^k||^2_2$
    \State Compute $\frac{\partial \mathcal{L}}{\partial \bm{\uptheta}_k}$
    \State $\bm{\uptheta}_{k+1} \gets \mathcal{O}\left( \bm{\uptheta}_{k}, l_r, \frac{\partial \mathcal{L}}{\partial \bm{\uptheta}_k} \right)$ \Comment{Optimizer step}
\EndFor

\State ${\hat{\mathbf{x}}}^\mathrm{R2R} \gets \mathbf{0}$ 
\For{$m \gets 1$ to $M$}\Comment{Monte-Carlo inference}
    \State $\mathbf{n}_s^m \gets \mathcal{N}(\mathbf{0}, \sigma \mathbf{I})$
    \State $\mathbf{z}_1^m = \mathbf{y} + \mathbf{D}^T \mathbf{n}_s^m$
    \State ${\hat{\mathbf{x}}}^\mathrm{R2R} \gets {\hat{\mathbf{x}}}^\mathrm{R2R} + A_{\bm{\uptheta}_K}(\mathbf{z}_1^m)$
\EndFor
\State ${\hat{\mathbf{x}}}^\mathrm{R2R} \gets \frac{1}{M} {\hat{\mathbf{x}}}^\mathrm{R2R}$
\end{algorithmic}
\end{algorithm}

%
%
%%% Algo NaC %%%%%
\begin{algorithm}[ht!]
\caption{Computing ${\hat{\mathbf{x}}}^\mathrm{NaC}$}\label{alg:nac}
\begin{algorithmic}
\Require \\
Denoising algorithm $A_{\bm{\uptheta}}$ \\
$\mathbf{y} = \mathbf{x} + \mathbf{n}$, $\mathbf{n}$ weak noise from distribution $\mathcal{D}$ \Comment{\textit{$\mathbf{n}$ does not need to be additive}} \\
Initial parameters $\bm{\uptheta}_0$ \\
Learning rate $l_r$ \\
Optimizer $\mathcal{O}$ \\
Number of iterations $K$ 

\For{$k \gets 1$ to $K$ }
    \State $\mathbf{n}_s^k \sim \mathcal{D}$
    \State $\mathbf{z}^k = \mathbf{y} + \mathbf{n}_s^k$ \Comment{can be done with other noise types}
    \State $\mathbf{u}_k \gets A_{\bm{\uptheta}_k}(\mathbf{z}^k)$
    \State $\mathcal{L} \gets ||\mathbf{u}_k - \mathbf{y}||^2_2$
    \State Compute $\frac{\partial \mathcal{L}}{\partial \bm{\uptheta}_k}$
    \State $\bm{\uptheta}_{k+1} \gets \mathcal{O}\left( \bm{\uptheta}_{k}, l_r, \frac{\partial \mathcal{L}}{\partial \bm{\uptheta}_k} \right)$ \Comment{Optimizer step}
\EndFor
\State ${\hat{\mathbf{x}}}^\mathrm{NaC} \gets A_{\bm{\uptheta}_K}(\mathbf{y}_1)$
\end{algorithmic}
\end{algorithm}
%
%
%%% Algo Nr2N %%%%%
\begin{algorithm}[t]
\caption{Computing ${\hat{\mathbf{x}}}^\mathrm{Nr2N}$}\label{alg:nr2n}
\begin{algorithmic}
\Require \\
Denoising algorithm $A_{\bm{\uptheta}}$ \\
$\mathbf{y} = \mathbf{x} + \mathbf{n}$, $\mathbf{n} \sim \mathcal{D}$ \Comment{For $\mathbf{n} \sim \mathcal{N}(\mathbf{0}, \sigma \mathbf{I})$, see Algorithm \ref{alg:nr2n_gaussian}}\\
Initial parameters $\bm{\uptheta}_0$ \\
Learning rate $l_r$ \\
Optimizer $\mathcal{O}$ \\
Number of iterations $K$ 

\For{$k \gets 1$ to $K$ }
    \State $\mathbf{n}_s^k \sim \mathcal{D}$
    \State $\mathbf{z}^k = \mathbf{y} + \mathbf{n}_s^k$
    \State $\mathbf{u}_k \gets A_{\bm{\uptheta}_k}(\mathbf{z}^k)$
    \State $\mathcal{L} \gets ||\mathbf{u}_k - \mathbf{y}||^2_2$
    \State Compute $\frac{\partial \mathcal{L}}{\partial \bm{\uptheta}_k}$
    \State $\bm{\uptheta}_{k+1} \gets \mathcal{O}\left( \bm{\uptheta}_{k}, l_r, \frac{\partial \mathcal{L}}{\partial \bm{\uptheta}_k} \right)$ \Comment{Optimizer step}
\EndFor

\State $\mathbf{n}_s \sim \mathcal{D}$
\State $\mathbf{z}^k = \mathbf{y} + \mathbf{n}_s$ 
\State ${\hat{\mathbf{x}}}^\mathrm{NaC} \gets 2 A_{\bm{\uptheta}_K}(\mathbf{z}) - \mathbf{z}$

\end{algorithmic}
\end{algorithm}

%%% Algo Nr2N  Gaussian Case%%%%%
\begin{algorithm}[t]
\caption{Computing ${\hat{\mathbf{x}}}^\mathrm{Nr2N}$ (Gaussian noise case)}\label{alg:nr2n_gaussian}
\begin{algorithmic}
\Require \\
Denoising algorithm $A_{\bm{\uptheta}}$ \\
$\mathbf{y} = \mathbf{x} + \mathbf{n}$, $\mathbf{n} \sim \mathcal{N}(0,\sigma)$ \\
Variance parameter $\alpha \in ]0,1]$\\
Initial parameters $\bm{\uptheta}_0$ \\
Learning rate $l_r$ \\
Optimizer $\mathcal{O}$ \\
Number of iterations $K$ 

\For{$k \gets 1$ to $K$ }
    \State $\mathbf{n}_s^k \gets \mathcal{N}(\mathbf{0}, \alpha \sigma \mathbf{I})$
    \State $\mathbf{z}^k = \mathbf{y} + \mathbf{n}_s^k$
    \State $\mathbf{u}_k \gets A_{\bm{\uptheta}_k}(\mathbf{z}^k)$
    \State $\mathcal{L} \gets ||\mathbf{u}_k - \mathbf{y}||^2_2$
    \State Compute $\frac{\partial \mathcal{L}}{\partial \bm{\uptheta}_k}$
    \State $\bm{\uptheta}_{k+1} \gets \mathcal{O}\left( \bm{\uptheta}_{k}, l_r, \frac{\partial \mathcal{L}}{\partial \bm{\uptheta}_k} \right)$ \Comment{Optimizer step}
\EndFor
\State $\mathbf{n}_s \sim \mathcal{D}$
\State $\mathbf{z}^k = \mathbf{y} + \mathbf{n}_s$
\State ${\hat{\mathbf{x}}}^\mathrm{NaC} \gets \frac{(1+ \alpha^2) A_{\bm{\uptheta}_K}(\mathbf{z}) - \mathbf{z}}{\alpha^2} $
\end{algorithmic}
\end{algorithm}

\clearpage
\printbibliography